\documentclass[12pt]{article}
\usepackage{a4wide}

\usepackage{graphicx} 
\usepackage{authblk}

\usepackage[breaklinks]{hyperref}
\usepackage{booktabs}

\begin{document}
\title{Probability distributions for quantum stress tensors in 
two and four dimensions}



\author[1]{Christopher J. Fewster}
\author[2]{L. H. Ford} 
\author[3]{Thomas A. Roman}

\affil[1]{Department of Mathematics, University of York,  
Heslington, York YO10 5DD, \newline  United Kingdom}
\affil[2]{Institute of Cosmology, Department of Physics and
  Astronomy, 
Tufts University, \newline Medford, Massachusetts 02155, USA }
\affil[3]{ 
Department of Mathematical Sciences, Central Connecticut State
University, \newline
New Britain, Connecticut 06050, USA}


\maketitle

\begin{abstract}
The probability distributions for the smeared energy densities of quantum fields, in the 
two and four-dimensional Minkowski vacuum are discussed. These distributions share the property that 
there is a lower bound at a finite negative value, but no upper bound. Thus arbitrarily large positive energy 
density fluctuations are possible. In two dimensions we are able to give an exact unique analytic 
form for the distribution. However, in four dimensions, we are not able to give closed form expressions
 for the probability distribution, but rather use calculations of a finite number of moments to estimate
the lower bound, and the asymptotic form of the tail of the distribution. 
The first 65 moments are used for these purposes. 
All of our four-dimensional results are subject to the caveat that
these distributions are not uniquely determined by the moments. 
One can apply the asymptotic form of the electromagnetic energy density 
distribution to estimate the nucleation
rates of black holes and of Boltzmann brains.
\end{abstract}

\section{Introduction}

There has been extensive work in recent decades on the definition and
use of the expectation value of a quantum stress tensor operator. 
However, the semiclassical theory does not describe the effects of
quantum fluctuations of the stress tensor around its expectation
value.

One way to examine these fluctuations is through the
probability distribution for individual measurements of a smeared stress tensor operator. This
distribution was given recently for Gaussian averaged
stress tensors operators in
two-dimensional flat spacetime~\cite{FFR10:FFRPrague} using analytical methods, 
and more recently for averaged stress tensors in
four-dimensional spacetime from calculations of a finite set of
moments. (Throughout our discussion, all quadratic operators
are understood to be normal-ordered with respect to the Minkowski
vacuum state.)

\subsection {Quantum Inequalities}

Quantum inequalities are lower bounds on the  {\it expectation
values} of the smeared energy density operator in arbitrary quantum 
states~\cite{F78:FFRPrague},~\cite{F91:FFRPrague},~\cite{FR95:FFRPrague},
~\cite{FR97:FFRPrague},~\cite{Flanagan97:FFRPrague},~\cite{FewsterEveson98:FFRPrague}.
If we sample in time along the worldline of an inertial observer, the quantum inequality takes the form
\begin{equation}
\int_{-\infty}^\infty f(t)\, \langle T_{\mu \nu} u^{\mu} u^{\nu} \rangle \, dt \geq
-\frac{C}{\tau^d} \,,
 \label{eq:QI}
\end{equation}
where $T_{\mu \nu} u^{\mu} u^{\nu}$ is the normal-ordered 
energy density operator, which is classically 
non-negative, $t$ is the observer's proper time, and $f(t)$ 
is a sampling function with characteristic width $\tau$.  
Here $C$ is a numerical constant, typically small
compared to unity, $d$ is the number of spacetime dimensions, 
and we work in units where $c=\hbar=1$.

Although quantum field theory allows negative expectation values of the energy
density, quantum inequalities place strong constraints on the effects of this negative
energy for violating the second law of thermodynamics~\cite{F78:FFRPrague}, maintaining
traversable wormholes~\cite{FR96:FFRPrague} or warpdrive spacetimes~\cite{PF97:FFRPrague}. 
The implication of Eq.~(\ref{eq:QI})
is that there is an inverse power relation between the magnitude and duration
of negative energy density. 

For a massless scalar field in two-dimensional spacetime, 
Flanagan~\cite{Flanagan97:FFRPrague}  has found a 
formula for the constant $C$ for a given $f(t)$ which makes Eq.~(\ref{eq:QI})
an optimal inequality. This formula is
\begin{equation}
C = \frac{1}{6 \pi} \, \int^\infty_{-\infty} du \left( \frac{d}{du} \sqrt{g(u)}\right)^2\,,
 \label{eq:Flanagan}
\end{equation}
where  $f(t)=\tau^{-1}g(u)$ and $u=t/\tau$.
In four-dimensional spacetime, 
Fewster and  Eveson~\cite{FewsterEveson98:FFRPrague}  have
derived an analogous formula for $C$, but in this case the bound is not
necessarily optimal. 

\section{Shifted Gamma Distributions - 2D Case}
\label{sec:SGD}

In two-dimensional Minkowski spacetime, we determined the
probability distribution for individual measurements, in the vacuum state, 
of the Gaussian sampled energy density to be 
\begin{equation}
\rho =\frac1{\sqrt{\pi}\,\tau} \int_{-\infty}^{\infty} 
T_{tt}\, {\rm e}^{-t^2/\tau^2} \, dt \,.
\end{equation}
This was achieved by
finding a closed form expression for the generating function of the moments $\langle \rho^n\rangle$
of $\rho$, from which the probability distribution was obtained. The definition of the $n$'th moment 
of the distribution of a variable $x$ is given by 
 \begin{equation}
a_n = \int x^n \, P(x)\, dx \,.   \label{eq:moment} 
\end{equation}

The resulting distribution is conveniently expressed in terms of  the dimensionless variable
$x = \rho \, \tau^2$ and is a shifted Gamma distribution:
\begin{equation}
P(x) =\vartheta(x+x_0)
\frac{\beta^{\alpha}(x+x_0)^{\alpha-1}}{\Gamma(\alpha)} 
\exp(-\beta(x+x_0)) \,,
\label{eq:shifted_Gamma}
\end{equation}
with parameters
\begin{equation}
x_0 = \frac{1}{12\pi},\qquad \alpha = \frac{1}{12}, \qquad
\beta = \pi \,.
\end{equation}
Here $x = -x_0$ is the lower bound of the distribution. 

\begin{figure} 
\begin{center}
\includegraphics[width=11cm]{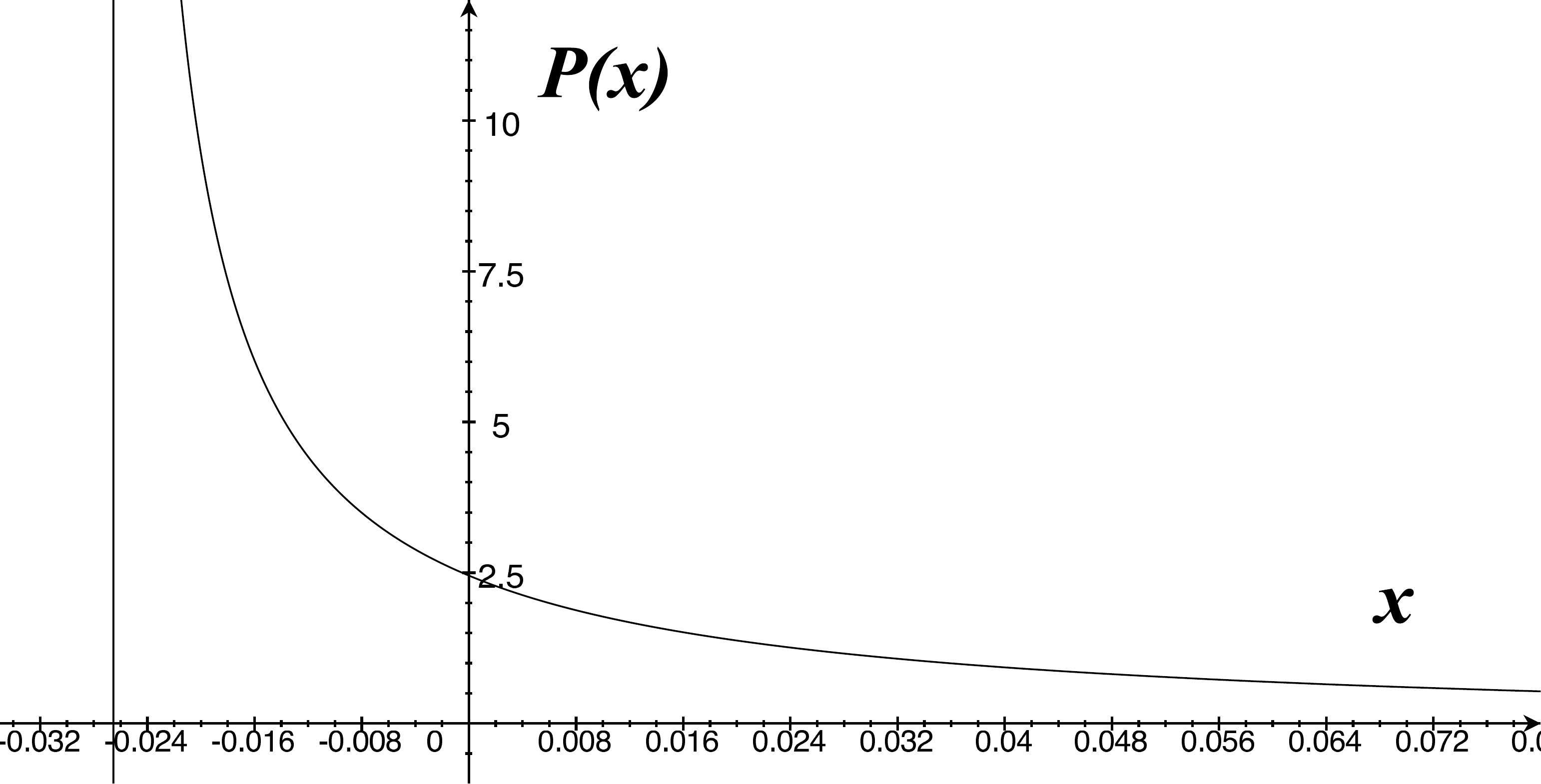}
\end{center}
\label{fig:2Dprob} 
\caption{The graph of $P(x)$ vs $x$ of the probability 
distribution function for the energy density, $\rho$, of a massless scalar field sampled
in time with a Gaussian of width $\tau$. Here $x = \rho \tau^2$.
The distribution has an integrable singularity at the optimal quantum inequality bound $x=- x_0= -1/12\pi$.} 
\end{figure}

The lower bound, $-x_0$, for the probability distribution for energy density
fluctuations in the vacuum is exactly Flanagan's optimum lower bound,
Eq.(~\ref{eq:Flanagan}), on
the Gaussian sampled expectation value. As was argued in 
Ref.~\cite{FFR10:FFRPrague},
this is a general feature, giving a deep connection between quantum inequality
bounds and stress tensor probability distributions. The quantum inequality
bound is the lowest eigenvalue of the sampled operator, and is hence
the lowest possible expectation value and the smallest result which
can be found in a measurement. That the probability distribution for
vacuum fluctuations actually extends down to this value is more
subtle and depends upon special properties of the vacuum state, and is implied by 
the Reeh-Schlieder theorem. 

There is no upper bound on  $P(x)$, as arbitrarily large values of the
energy density can arise in vacuum fluctuations. Nonetheless, for the
massless scalar field, negative values are much more likely; 84\% of
the time, a measurement of the Gaussian averaged energy density will
produce a negative value. However, the positive values found the
remaining 16\% of the time will typically be much larger, and the
average first moment of $P(x)$ will be zero. 

Furthermore, the probability distribution for the two-dimensional stress
tensor is uniquely determined by its moments, as a
consequence of the Hamburger moment theorem
~\cite{Simon:FFRPrague}.  
This condition is a sufficient, although not necessary, condition
for uniqueness, and is fulfilled by the moments of the shifted Gamma
distribution. 

\section{The  4D Case}
\label{sec:4D}

In four dimensions, the
operators $\rho_S$, and $\rho_{EM}$ all have
dimensions of $length^{-4}$. Their probability distributions $P(x)$ are  
taken to be functions of the dimensionless variable 
\begin{equation}
x = (4\pi \, \tau^2)^2 \, A\,,
\end{equation}
where $A$ is the Lorentzian time average of  $\rho_S$, and $\rho_{EM}$, 
where $\rho_S$ and $\rho_{EM}$ are the smeared energy density operators for 
the massless scalar field, and electromagnetic fields, respectively.

The distributions were calculated numerically from 65 moments~\cite{FFR12:FFRPrague} 
The situation here is less straightforward. In this case, the
moments grow too rapidly to satisfy the Hamburger moment 
criterion. Unfortunately, this means that we cannot be guaranteed
of finding  a unique probability distribution $P(x)$ from these moments. These probability 
distributions share some of the main characteristics of their two-dimensional counterparts. 
They have a lower bound but no upper bound. Our  
our techniques allow us to give approximate lower bounds and the 
asymptotic forms of the tails of the distributions.

Our estimates for the lower bounds are 
\begin{equation}
-x_0(\rho_{EM})   \approx  -0.0472 \qquad  
-x_0(\rho_S)   \approx -0.0236\,.
 \label{eq:bounds2}
\end{equation}
These are also estimates of the optimal quantum inequality bounds
for each field. In contrast, the non-optimal bound for $\rho_S$, given by 
the method of Fewster and Eveson~\cite{FewsterEveson98:FFRPrague}, 
is $-x_0(FE) =- 27/128 \approx -0.21$, which is an order of magnitude larger. 

It is of interest to note that the magnitudes of the dimensionless lower bounds, given in
Eq.~(\ref{eq:bounds2}) are small compared to unity. 
The fact that the
probability distribution has a long positive tail, and must have a unit zeroth
moment and a vanishing first moment, implies that the total probability
of a negative value to be substantial.
%
The small magnitudes of $x_0(\rho_{S})$
and  $x_0(\rho_{EM})$ imply strong constraints on the magnitude of negative energy which
can arise either as an expectation value in an arbitrary state, or as a fluctuation in the  
vacuum. They also imply that an individual measurement of the sampled energy density in the 
vacuum state is very likely to yield a negative value. 

One can show that the asymptotic behavior of the tail of the probability distribution 
is determined by the moments, even if the exact probability distribution is not uniquely 
determined. Our fitted tail decreases asymptotically as 
\begin{equation}
P_{{\rm fit}} \sim e^{-a x^{1/3} } \,,
\end{equation}
where $a$ is a constant. 
We are also able to show that no distribution with the same moments can have a tail 
which decreases at a faster rate than ours. 

By contrast, the tail of a Boltzmann distribution for thermal fluctuations falls off as 
\begin{equation}
P_{{\rm  Boltzmann}} \sim e^{-\beta x} \,,
\end{equation}
where $\beta$ is a constant. Therefore vacuum fluctuations outweigh thermal 
fluctuations at high energies.

\begin{figure} 
\begin{center}
\includegraphics[width=12cm]{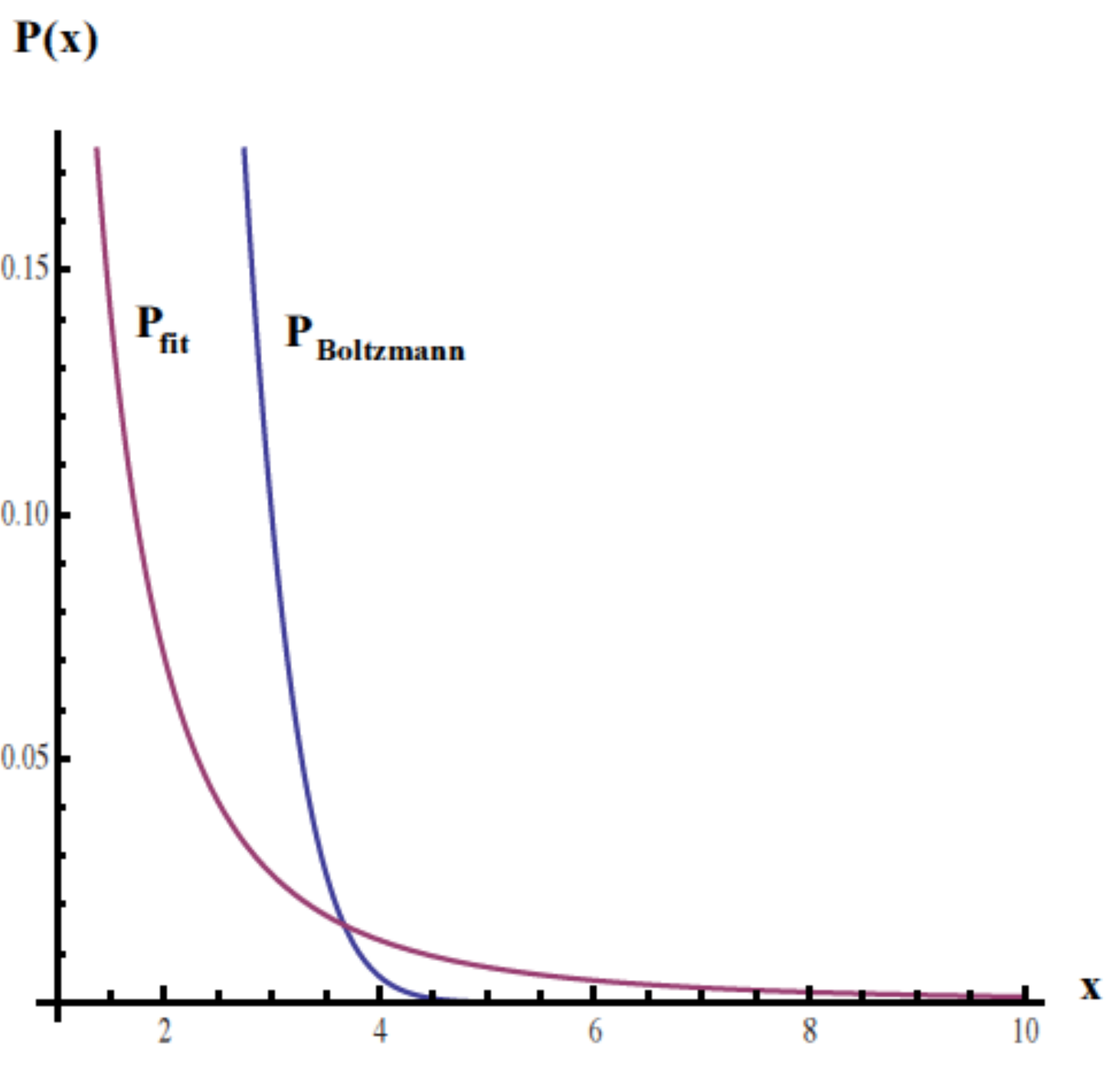}
\end{center}
\label{fig:Ptail-vs-Boltzmann} 
\caption{The figure shows a comparison of the asymptotic form of the tails of both our fitted 
distribution for vacuum fluctuations and for the thermal fluctuations described by the 
Boltzmann distribution. At high energies, vacuum fluctuations outweigh thermal fluctuations.} 
\end{figure}

\subsection{Application: Black Hole Nucleation}

The fact that the energy density probability distribution has 
a long positive tail implies a finite probability for the 
nucleation of black holes out of the Minkowski vacuum via 
large, though infrequent positive fluctuations (see
Ref~\cite{FFR12:FFRPrague}).
This probability cannot be too large, of course, or it will 
conflict with observation. Our estimate of the probability depends only on the 
asymptotic form of the tail. (One can use similar arguments to estimate the probability 
of ``Boltzmann brains''~\cite{Bbrains:FFRPrague} nucleating out of the vacuum.)

\section{Summary} 

We have found that the probability distribution for vacuum fluctuations of the 
Gaussian-smeared energy density for a massless scalar field 
in two-dimensional spacetime is {\it uniquely} 
defined by a shifted gamma distribution. The distribution has a negative lower bound but 
no upper bound. It has an integrable singularity (i.e., a ``spike") 
at the lower bound. In addition, we find that there is a deep connection between the lower 
bound of the distribution and the quantum inequalities. In fact  the lower bound 
of the distribution coincides {\it exactly} with the {\it optimal} quantum inequality 
bound for a Gaussian sampling function, derived earlier by Flanagan. 

The lower bound is 
very small in magnitude, but the probability distribution is large in the region between zero 
and the lower bound. As a result, rather surprisingly, the 
probability of obtaining a negative result in an 
individual measurement is $84\%$! Although the negative fluctuations are very frequent, 
they are small in magnitude. As a result, one would not expect to see large effects of 
negative energy (e.g., violations of the second law, wormholes, warpdrives, etc.) 
nucleating out of the vacuum. However, the distribution has a long positive tail, which
guarantees that the frequent but small negative energy density 
fluctuations are balanced by the much rarer but larger positive energy fluctuations. Therefore, 
the expectation value of the energy density in the Minkowski vacuum state is zero. 
It is quite remarkable that the quantum inequalities which are 
bounds on the {\it expectation value} of the energy density in 
an {\it arbitrary} quantum state, should be so intimately related to the probability 
distribution of {\it individual} measurements of the energy density made in the 
{\it vacuum} state.

In four dimensions, we find similarities with the two-dimensional  case, in that there is a 
lower bound but no upper bound. We are able to give numerical estimates of the lower 
bounds, i.e., the optimal bounds, and the asymptotic form of the tails. 
The lower bounds are negative with small 
magnitudes. However, our methods do not allow us to determine whether there is a 
``spike'' at the lower bound, as in two dimensions. Nonetheless, the low magnitudes 
of the lower bounds indicate that a significant fraction of the probability must lie in the 
negative region. Therefore, as in the two-dimensional case, the probability of obtaining 
a negative value in an individual measurement is quite high. The long positive tail 
drops off more slowly than that of a Boltzmann distribution, which implies that 
vacuum fluctuations dominate over thermal fluctuations at high energies. 

Unfortunately, it seems likely that it is not possible to uniquely determine the 
four-dimensional distributions from the moments alone, as the latter do not obey 
the Hamburger moment condition. Nonetheless, we are able to glean some 
information from the moments. For example, we can determine that no distribution 
with the same moments as ours can have a tail which decreases faster than ours. 
The asymptotic forms of the long positive tail allow us to estimate the probability 
of nucleation of (small) black holes and ``Boltzmann brains" out of the vacuum.

Clearly further work can be done on this subject. One topic would be to see what 
additional information can be obtained from our calculated four-dimensional 
probability distributions, even if they cannot be uniquely determined from their moments. 
For example, does the ``spike'' behavior persist in four dimensions as well as in two, and 
what is its physical significance? Another would be to determine what  the optimal 
quantum inequality bounds actually are.  It would 
also be useful to try various sampling functions. Can 
the probability distributions and optimal bounds can be obtained by 
other methods which do not have the limitation of the ambiguities in the 
moment methods? There is more to do to explore the physical content 
of stress-tensor fluctuations.


\end{document}